\documentclass[draft]{agujournal2019}
\usepackage{soul}
\usepackage[finalnew]{trackchanges} %for better track changes. finalnew option will compile document with changes incorporated.
\usepackage{url} %this package should fix any errors with URLs in refs.
\usepackage{amsmath}
\usepackage{tabularx}
\usepackage{gensymb}
\usepackage{amssymb}
\usepackage{array}

\soulregister\ref7
\soulregister\cite7
\soulregister\citeA7
\soulregister\citeNP7
\soulregister\url7

\draftfalse

\journalname{JGR: Planets}

\addeditor{eryn}
\begin{document}

	%% ------------------------------------------------------------------------ %%
	%  Title
	%
	% (A title should be specific, informative, and brief. Use
	% abbreviations only if they are defined in the abstract. Titles that
	% start with general keywords then specific terms are optimized in
	% searches)
	%
	%% ------------------------------------------------------------------------ %%

	\title{Higher Martian atmospheric temperatures at all altitudes \change{lead to enhanced}{increase the} D/H fractionation\add{ factor} and water loss}
	%% ------------------------------------------------------------------------ %%
	%
	%  AUTHORS AND AFFILIATIONS
	%
	%% ------------------------------------------------------------------------ %%

	\authors{E. M. Cangi\affil{1,2}, M. S. Chaffin\affil{1}, J. Deighan\affil{1}}

	\affiliation{1}{Laboratory for Atmospheric and Space Physics}
	\affiliation{2}{University of Colorado Boulder}

	\affiliation{1}{3665 Discovery Dr, Boulder, CO 80303}
	\affiliation{2}{Boulder, CO}
	%(repeat as many times as is necessary)

	%% Corresponding Author:
	% Corresponding author mailing address and e-mail address:

	\correspondingauthor{Eryn Cangi}{eryn.cangi@colorado.edu}

	%% Keypoints, final entry on title page.

	%  List up to three key points (at least one is required)
	%  Key Points summarize the main points and conclusions of the article
	%  Each must be 100 characters or less with no special characters or punctuation and must be complete sentences

	\begin{keypoints}
	\item The fractionation factor $f$ \change{ranges from $10^{-5}$ to $10^{-1}$ for thermal escape only, and 0.03 to 0.1 for thermal + non-thermal escape.}{is most strongly controlled by non-thermal escape of atomic deuterium (D).}
	\item \add{Larger} $f$ \change{is insensitive to}{correlates minimally with higher} atmospheric temperature at the surface, but \change{depends on}strongly \add{with }exobase and tropopause temperatures.
	\item Using our results for $f$, we calculate total water lost from Mars to be between 66-\change{123}{122} m GEL, which is likely a lower bound.
	\end{keypoints}

	%% ------------------------------------------------------------------------ %%
	%
	%  ABSTRACT and PLAIN LANGUAGE SUMMARY
    ------------------------------------------------------------------------ %%

\begin{abstract}
Much of the water that once flowed on the surface of Mars was lost to space long ago, and the total amount lost remains unknown. Clues to the amount lost can be found by studying hydrogen (H) and its isotope deuterium (D), \remove{both of }which are produced when atmospheric water molecules H$_2$O and HDO dissociate. The \add{difference in escape efficiencies of} \remove{freed }H and D \remove{atoms then escape to space at different rates due to their different masses, }\change{leaving}{(which leads to} an enhanced D/H ratio\change{. The rate of change of D/H}{)} is referred to as the fractionation factor $f$. Both the D/H ratio and $f$ are necessary to estimate water loss; thus, if we can constrain the range of $f$ \add{and understand what controls it}, we will be able to estimate water loss more accurately. In this study, we use a 1D photochemical model of the \add{neutral} Martian atmosphere to determine how $f$ depends on assumed temperature and water vapor profiles. We find that \change{for most Martian atmospheric conditions, $f$ varies between $10^{-1}$ and $10^{-5}$;}{the exobase temperature most strongly controls the value of $f$ for thermal escape processes. When we include estimates of non-thermal escape from other studies, we find that the tropopause temperature is also important. Overall,} for the standard Martian atmosphere, $f=0.002$ for thermal escape\remove{ processes}, and $f=0.06$ \change{when}{for} \remove{both} thermal \change{and}{+} non-thermal escape\remove{ are considered}. \remove{Using these results, }\change{w}{W}e estimate that Mars has lost at minimum 66-\change{123}{122} m GEL of water. \add{Importantly, }\change{O}{o}ur results demonstrate that the value of $f$ \change{is almost completely controlled by}{depends critically on} non-thermal escape of D, and that \remove{photochemical }modeling studies that include \add{D/H }fractionation must \remove{thus }model both neutral and ion processes throughout the atmosphere.

\end{abstract}

\section*{Plain Language Summary}

\change{Much of the water that once flowed on the surface of Mars was lost to space long ago, and the total amount lost remains unknown. Clues can be found by studying the two types of water: the familiar H$_2$O, and HDO, a heavier version of water.}{Mars used to have lots of water, but has lost it over time.} When water molecules break apart in the atmosphere, they release hydrogen (H) and its heavier twin deuterium (D), which escape to space at different rates, removing water from Mars. The \remove{difference in }escape efficiency \change{between}{of} \change{H and D}{D compared to H} is called the fractionation factor $f$. The goal of this study is two-fold: to understand how $f$ varies with different atmospheric conditions and the processes that control it, and to use that information to estimate water loss from Mars. To do this, we model \change{the atmospheric chemistry of Mars, testing}{the Martian atmosphere to test how} different atmospheric temperatures and water vapor content \remove{to understand how they }affect $f$. \add{We find that the most important thing affecting $f$ is loss of D via processes involving interaction with planetary ions or the solar wind, rather than loss of D that is hot enough to exceed escape velocity. This implies future studies must include ion chemistry to accurately calculate $f$. We also find that generally, temperatures above 100 km strongly affect the value of $f$.} Using the\add{se} results\remove{ for $f$}, we calculate that Mars has lost enough water to cover the whole planet in a layer between 66-\change{123}{122} m deep\remove{, in agreement with other photochemical modeling studies, but still short of geological estimates}.

\section{The D/H Fractionation Factor and Loss of Martian Water to Space}

The surface of Mars is marked with ample evidence of its wetter past. Today, water on Mars exists only in the polar caps, subsurface ice, and atmosphere, but geomorphological and geochemical evidence points to significant alteration of the surface by liquid water. The presence of compounds like jarosite and hematite indicate past pooling and evaporation \cite{Squyres2004, Klingelhofer2004}, while substantial evidence of hydrated silicates supports the theory that ancient river deltas, lake beds, catastrophic flood channels, and dendritic valley networks were formed by water \cite[and references therein]{Carr2010, Ehlmann2014}. Because the contemporary Martian climate \change{is too cold and too low-pressure to }{cannot }support liquid water on the surface, \remove{all this evidence means that }Mars must have \add{once} had \remove{both }a thicker and warmer atmosphere\remove{, and therefore a stronger greenhouse effect. Identifying the greenhouse gas responsible is the topic of ongoing studies \cite{Ramirez2014, Wordsworth2017}}. \remove{Regardless, }\change{t}{T}he Mars science community generally agrees that \add{the atmosphere has escaped over time, with }a significant amount \change{of the once-thick Martian atmosphere has escaped to space over time. Most of this escape occurs} in the form of thermal escape of H, in which a fraction of H atoms are hot enough that their velocity exceeds the escape velocity. Because H is primarily found in water on Mars, \change{integrated atmospheric escape}{this} has effectively desiccated the planet \cite{Jakosky2018}.

A significant indicator of this loss of water to space is the elevated D (deuterium, $^2$H or D) to H (hydrogen, $^1$H) ratio, which we will abbreviate as $R_{dh}$. \change{On Mars, }{Because }water (either as H$_2$O or HDO) is the primary reservoir of both H and D\change{. W}{, w}hen we talk about the D/H ratio, we are thus usually referring to \remove{the }D/H\remove{ ratio as} measured in water:

\begin{equation}
R_{dh} = \frac{\text{D in HDO}}{\text{H from HDO}+\text{H from H$_2$O}} = \frac{[HDO]}{[HDO] + 2[H_2O]} \approxeq \frac{[HDO]}{2[H_2O]}
\end{equation}

Here, $[X]$ represents a molecule's abundance; H sourced from HDO is negligible compared to H sourced from H$_2$O. This ratio evolves according to differential escape of D and H; D, being twice as massive as H, is less likely to escape. This difference can be characterized as a relative efficiency, the fractionation factor $f$:

\begin{equation} \label{eq:f}
f = \frac{\phi_{\text{D}} / \phi_{\text{H}}}{[HDO]_s/2[H_2O]_s} = \frac{\phi_{\text{D}} / \phi_{\text{H}}}{{R_{dh,s}}}
\end{equation}

\noindent where $\phi$ represents \remove{outgoing }fluxes to space, and the $s$ subscript specifies the near-surface atmospheric reservoir, which approximates the total amount in the atmosphere. As it represents efficiency of D escape, $f$ takes on values between 0 and 1. When $f$ is 0, D is completely retained on the planet, and cumulative water loss must have been lower than for $f \neq 0$. When $f=1$, the \textit{ratio} of escaping to retained atoms is the same for both D and H, and there is no mass effect on the escape rates. In this scenario, no amount of escape is sufficient to change the D/H ratio in any species. In practice, $f$ is somewhere in between these extremes.

Over geologic time, this fractionation manifests as an enhancement of the D/H ratio compared to the Earth ratio of $1.6\times 10^{-4}$ \cite{Yung1988}, called SMOW (for the measured source, Standard Mean Ocean Water). A planet's D/H ratio is often quoted as a multiple of the Earth value. At present, multiple measurements put the global mean $R_{dh}$ on Mars between 4 and 6 $\times$ SMOW \cite{Owen1988, Bjoraker1989, Kras1997, Encrenaz2018, Vandaele2019}, with some variations occurring on local spatial and temporal scales \cite{Villanueva2015, Clarke2017, Encrenaz2018, Clarke2019, Villanueva2019}. This is most commonly interpreted as evidence for significant escape to space of H.

\add{To estimate the integrated amount of water lost, one can use }\change{C}{c}urrent estimates of the Martian water inventory, $R_{dh}$, and $f$ \remove{are used }with the Rayleigh distillation equation \remove{to estimate the integrated amount of water lost from Mars. The Rayleigh distillation equation }for H\remove{ on Mars is} \cite{Yung1998}:

\begin{equation} \label{eq:genRD}
R_{dh}(t) = R_{dh}(t=0) \left( \frac{[H](0)}{[H](t)} \right)^{1-f},
\end{equation}

\noindent \change{W}{w}here $t=0$ can be arbitrarily chosen. Because we use $R_{dh}$\change{,}{ and because water is the primary reservoir of H on Mars,} [H] is \change{a proxy for}{commonly replaced with} total water $W$ ($W$ = [H$_2$O] + [HDO]). Then $W(0)$, the total water on Mars at some point in the past $t=0$, is the sum of the water budget at time $t$ and the total water lost: $ W(0) = W(t)$ + $W_{text{lost}}$. Substituting $W$ for [H] and rearranging equation \ref{eq:genRD}, we obtain an expression for water lost from Mars:

\begin{equation} \label{eq:rd}
%\frac{[H_2O]_{i}}{[H_2O]_{c}} &= \left( \frac{(D/H)_{c}}{(D/H)_{i}} \right)^{\frac{1}{1-f}} \nonumber \\
%\frac{[H_2O]_{c}+ [H_2O]_{L}}{[H_2O]_c} &= \left( \frac{(D/H)_{c}}{(D/H)_{i}} \right)^{\frac{1}{1-f}} \nonumber \\
%\frac{[H_2O]_{c}+ [H_2O]_{L}}{[H_2O]_{c}} &= \left( \frac{(D/H)_c}{(D/H)_i} \right)^{\frac{1}{1-f}} \nonumber \\
W_{\text{lost}} = W(t) \left( \left( \frac{R_{dh}(t)}{R_{dh}(0)} \right)^{1/(1-f)} - 1\right)
\end{equation}

Most of the inputs to Equation \ref{eq:rd} are well-described. The current D/H ratio of exchangeable water (the atmosphere, seasonal polar caps, ground ice, and water adsorbed in the regolith), $R_{dh} (t)$, is $4-6\times$ SMOW as mentioned (we use 5.5 in this study). $R_{dh}(0)$ is usually taken to be that at Mars' formation, when it would have been similar to the Earth's D/H ratio \cite{Geiss1981}; $R_{dh}$ at other points in time can be obtained from \remove{analysis of Martian surface material. These studies are limited; }meteorite samples \cite[e.g.]{Usui2012} \change{provide some data, and }{or }in-situ analysis\remove{ at Mars more} \cite[e.g.]{Mahaffy2015}. The current water inventory in exchangeable reservoirs, $W(t)$, is estimated to be between 20-30 m GEL (global equivalent layer), the depth of water if the entire exchangeable inventory were rained onto the surface \annote{}{added Z \& P citations}\cite{Zuber1998, Plaut2007, Lasue2013, Villanueva2015, Carr2019}.

Prior studies \change{produced best estimates of}{estimated} the fractionation factor $f$, but its range of values under all plausible scenarios has been largely unexplored. \citeA{Yung1988} used a 1D photochemical model to calculate a first value of $f=0.32$ which has been frequently referenced in the years since. They explored the effects of certain chemical reactions on $f$, but did not test other parameters. \citeA{Kras1998} obtained $f=0.02$ by combining Hubble Space Telescope observations with a radiative transfer and 1D photochemical model. Later, \citeA{Kras2000} followed up with another study that tested the effects of two different models of eddy diffusion, finding values of $f=0.135$ and $f=0.016$. Two years later, \citeA{Kras2002} \remove{released another study that }found 3 values for $f$, depending on whether the solar cycle was at minimum ($f=0.055$), maximum ($f=0.167$), or mean ($f=0.082$), represented in the model by variation of the exobase temperature and non-thermal escape flux. Our goal is to advance this body of work by performing the first systematic parameter-space study of the fractionation factor with respect to the assumed atmospheric temperature and water vapor profiles.

\section{Building Our 1D Photochemical Model} \label{sect:model}

To best capture the mean behavior of the Martian atmosphere over long time scales, we use a 1D photochemical model, extended from the original developed by \citeA{Chaffin2017} to include D chemistry. The model uses standard photochemical techniques described in other studies \cite{Kras1993, Nair1994, Chaffin2017}, with the addition of the D-bearing species D, HD, HDO, OD, HDO$_2$, DO$_2$, and DOCO. The chemical reactions for D-bearing species came from several sources, including past papers \cite{Yung1988, Yung1989, Cazaux2010, Deighan2012}, NASA publications \cite{Sander2011}, and online databases \cite{NIST, KIDA, UMIST}. The full list of chemical reactions and reaction rates, as well as \change{information}{details} on \add{implemented} photochemical cross sections \add{\cite{Barfield1972, Nee1984, Cheng1999, Cheng2004}} and diffusion coefficients\add{ \cite{BanksNKockarts}}, is given in the Supporting Information. Photodissociation is driven by solar UV irradiation data from SORCE/SOLSTICE and TIMED/SEE \cite{LISIRD}, appropriate for solar mean conditions and scaled to Mars' orbit. For our primary input, we construct temperature and water vapor profiles designed to represent end-member states of the atmosphere, such that we fully constrain the range of plausible fractionation factor values.

A run of the model consists of the following steps: (1) loading the temperature and water vapor profiles, (2) establishing an initial condition of species number densities, (3) establishing boundary conditions (Table S3), \add{and }(4) stepping forward over 10 million years of simulation time until the atmosphere reaches chemical equilibrium, which is achieved when the combined escape flux of atomic H and D ($\phi_{\text{H}} + \phi_{\text{D}}$) is twice that of the escape flux of atomic O ($\phi_O$). \add{(This stoichiometric balance is required because H and D are primarily sourced from water.)} The model output comprises species number densities by altitude. By multiplying the \change{H and D}{H- and D-bearing atomic and molecular species} densities by their thermal effusion velocities \cite{Hunten1973}, we can calculate \remove{the escape fluxes of H and D, }$\phi_{\text{H}}$ and $\phi_{\text{D}}$\change{.}{:}

\begin{gather} \label{eq:escfluxH}
\phi_{\text{H}} = n_{\text{H}}v_{\text{H}} + 2n_{\text{H}_2}v_{\text{H}_2} + n_{\text{HD}}v_{\text{HD}} \\
\label{eq:escfluxD} \phi_{\text{D}} = n_{\text{D}}v_{\text{D}} + n_{\text{HD}}v_{\text{HD}}
\end{gather}

\noindent These fluxes are then used to calculate $f$ according to equation \ref{eq:f}.

A limitation of our model is that we do not include a full ionosphere. Instead, we approximate it by including a static profile of CO$_2^+$ \cite{Matta2013}, enabling the primary H-producing ion reaction in the Martian atmosphere; a similar tactic was used by \citeA{Yung1988}. Without a full ionosphere, we are not able to model non-thermal escape of H or D, as most non-thermal processes depend on ions. In an effort to estimate the relative importance of non-thermal processes to the fractionation factor, we estimate non-thermal effusion velocities for our model conditions, scaled from \change{\citeA{Kras2010}}{\citeA{Kras2002}}, described further in Section \ref{sect:results}.

\subsection{Reproductions of Past Studies} \label{sect:repro}

Before proceeding with our study, we attempted to reproduce the results by \citeA{Yung1988} and \citeA{Kras2002}. Their original results and our reproductions are shown in Figure S3. We achieved very good agreement with the results by \citeA{Yung1988}, finding $f=0.26$ versus their $f=0.32$\change{, with t}{. T}he small difference \change{being due to an inability to reproduce the exact same photodissociation rates due to self-consistent calculation}{is due to the only two differences between the model by \citeA{Yung1988} and ours. First, \citeA{Yung1988} manually fix their photodissociation rates, whereas our model calculates them. Second, they adopt the water profile used by \citeA{Liu1976}, fixing it below 80 km and allowing it to vary above, whereas we fix it throughout the atmosphere.} Our results for $f$ were consistent with \citeA{Kras2002} for solar maximum, but comparatively low for solar mean and minimum. We expect that this is because their model includes an ionosphere, allowing them to model non-thermal escape\remove{ of D}. To account for this, we added their results for non-thermal escape of D to our results for thermal escape, resulting in a slight overestimate\change{ of $f$ for all solar states}{ instead of an underestimate}. This change was a first hint at the importance of non-thermal escape to $f$. The remaining discrepancy is due to other significant model differences\add{, irreconcilable without rewriting our model}; for example, their model atmosphere has its lower bound at 80 km, while ours is at the surface. \add{This difference in model extent means that our two models also have significant differences in temperature structure and boundary conditions.}

\subsection{Model input: Temperature and Water Vapor Profiles} \label{sect:inputs}

Our temperature and water vapor vertical profiles remain fixed for the duration of a simulation. This allows us to examine the mean behavior of the atmosphere over long time scales.

\subsubsection{Temperature Profiles}

The piecewise temperature profile \add{is modeled with equation \ref{eq:temp}. The general form is based on measurements by the Viking orbiters \cite{Seiff1982}, and similar models have been used in other studies (e.g. \citeNP{McElroy1977, Nair1994, Kras2010}). In the lower atmosphere, the temperature decreases with altitude according to the dry adiabatic lapse rate $\Gamma$. At the tropopause (altitude $z_t$), temperature reaches a minimum and remains isothermal in the mesosphere. Above the mesopause (120 km), upper atmospheric density is low enough that UV heating is efficient, causing the temperature to increase rapidly with altitude.}

\begin{equation} \label{eq:temp}
T = \begin{cases}
T_{\text{exo}} - (T_{\text{exo}} - T_{\text{tropo}})\exp{\left(-\frac{(z-120)^2)}{(8T_{\text{exo}})}\right)} & z > 120 \text{ km} \\
T_{\text{tropo}} & z_{t} < z < 120 \text{ km} \\
T_{\text{surf}} + \Gamma z & z < z_{t}
\end{cases}
\end{equation}

We constrain this modeled profile with the temperature\add{s} at the surface ($T_{\text{surf}}$), tropopause ($T_{\text{tropo}}$), and exobase ($T_{\text{exo}}$). Constraining the temperature at these three points requires either $\Gamma$ or $z_t$ to vary; if they are both fixed, the profile will be over-constrained and discontinuous. We allow $z_t$ to vary \change{because}{as} it does \remove{vary }in reality\add{ (e.g., \citeNP{Forget2009})}; exactly what sets its altitude is less well defined than the dynamics of gas and dust, on which $\Gamma$ depends. We use $\Gamma = -1.4$ K/km, which is slightly lower than the standard dry adiabatic lapse rate due to warming effects from suspended dust \cite{Zahnle2008}.

\begin{figure}
	\centering
	\includegraphics[width=1\linewidth]{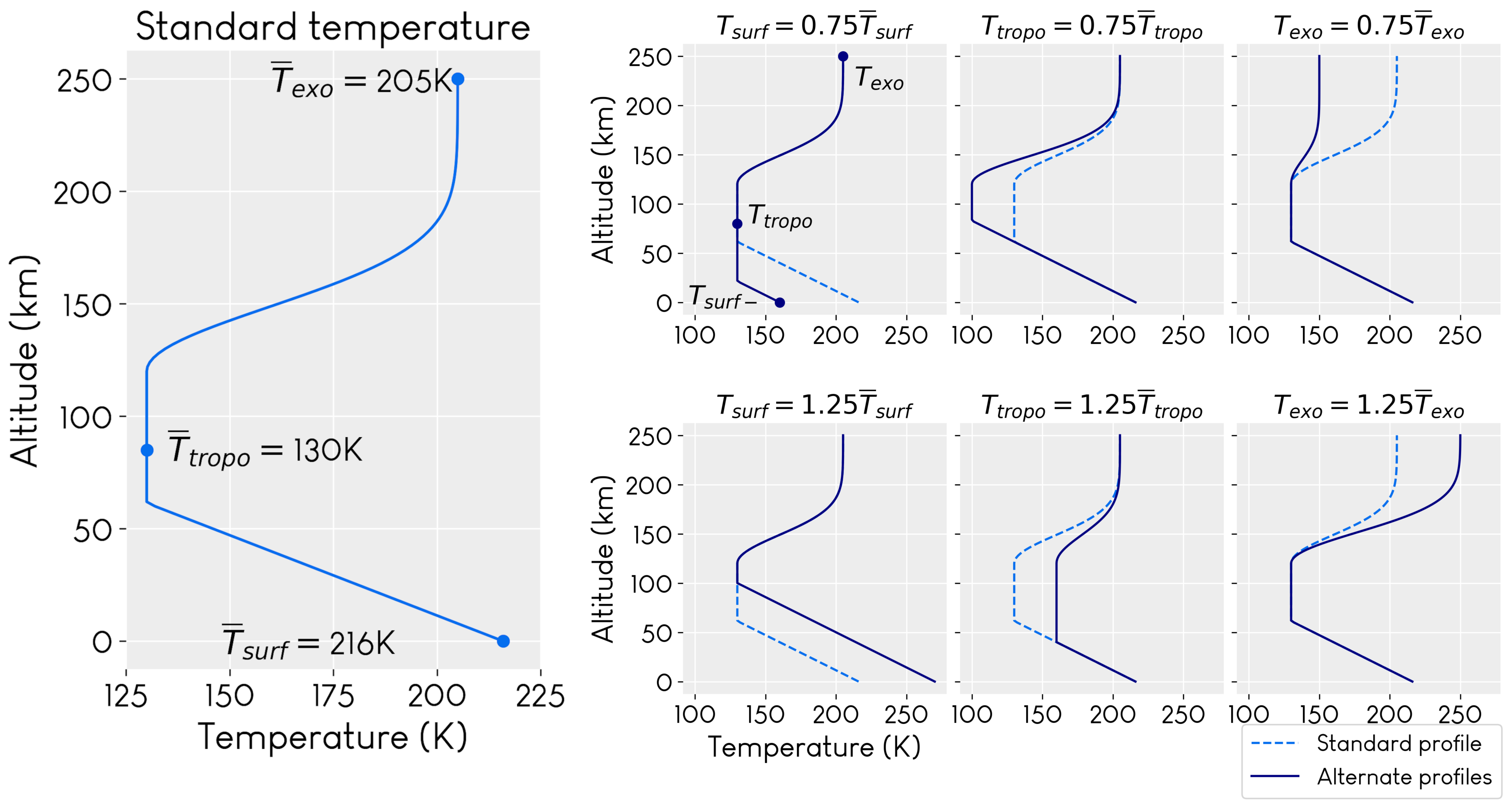}
	\caption{a) Our standard temperature profile used in the model, and b) alternate temperature profiles representing plausible climate extrema due to obliquity variations. Profiles are created by modifying the standard temperatures $\overline{T}_{\text{surf}}$, $\overline{T}_{\text{tropo}}$, or \textbf{$\overline{T}_{\text{exo}}$ }by $\pm25$\%. We do not consider effects of CO$_2$ condensation for cold temperatures, although this \change{is likely to}{would} be important in real\change{ity}{ atmospheres}. These profiles, along with the standard profile, are used to obtain the results in Figure \ref{fig:f-results-plot}. Table S4 gives specific values for $T_{\text{surf}}$, $T_{\text{tropo}}$. $T_{\text{exo}}$.}
	\label{fig:temp_profiles}
\end{figure}

For the first part of the study, we constructed a standard temperature profile representing current conditions on Mars, as well as 6 alternate profiles intended to represent plausible climate extrema driven by changing planetary obliquity throughout the last $\sim$10 million years of Mars' history, the maximum time over which \change{evolution of the obliquity}{obliquity evolution} can be analytically predicted. (On longer time scales, the obliquity evolves chaotically, making precise definition of climate parameters impossible \cite{Laskar2004}.) We used the Mars Climate Database (MCD) \cite{MCD} to obtain values for $T_{\text{surf}}$ ($z=0$), $T_{\text{tropo}}$ ($z=100$ km), and $T_{\text{exo}}$ ($z=250$ km) for different times of sol (local times 03:00, 09:00, 15:00, 21:00), Mars latitude (90$\degree$N, 45$\degree$N, 0$\degree$, 45$\degree$S, 90$\degree$S), and L$_s$ (90$\degree$ and 270$\degree$)\change{. T}{, then compared t}he mean temperatures across each of these parameters \remove{were then compared }with data from multiple missions to ensure consistency. The surface temperature was compared with the Curiosity Rover \cite{Vasavada2016, Audouard2016, Savijarvi2019}, Mars Global Surveyor Thermal Emission Spectrometer (TES) \cite{Smith2004}, and the Spirit/Opportunity Rovers' Mini-TES \cite{Smith2006}; the exobase temperature was compared with MAVEN data from multiple instruments \cite{Bougher2017, Stone2018, Thiemann2018}. The mean temperatures formed the standard profile, shown in Figure \ref{fig:temp_profiles}a. \change{The 6 alternate profiles are shown in Figure \ref{fig:temp_profiles}b. For each,}{Figure \ref{fig:temp_profiles}b shows the 6 alternate profiles, in which} we \change{either increased or decreased}{varied} one of $T_{\text{surf}}$, $T_{\text{tropo}}$, or $T_{\text{exo}}$ by \add{$\pm$}25\% of the standard value. This variation covers most values observed by current missions, as well as temperatures calculated \cite{Wordsworth2015} for obliquities of $\sim$25-45\degree predicted for the last 10 million years \cite{Laskar2004}. A table with the control temperatures for each profile is available in the Supporting Information. Together, the standard and alternate temperature profiles represent end-member cases for the Martian atmosphere.

\begin{figure}
	\centering
	\includegraphics[width=1\linewidth]{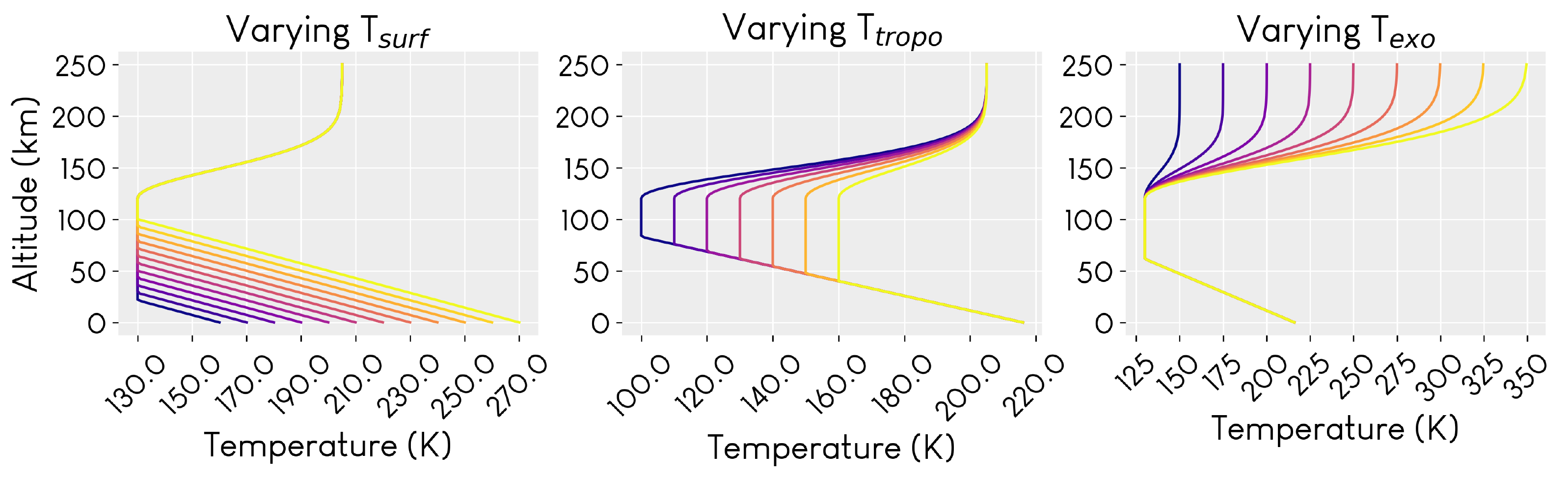}
	\caption{The full range of temperature profiles tested. Each panel represents a set of profiles in which one of the specifiable temperatures was varied. Results from the simulations using these profiles are shown in Figure \ref{fig:f-vs-temps}. Each color represents a different profile.}
	\label{fig:alltemps}
\end{figure}

In addition to these select profiles, we also created a larger set of temperature profiles with finer variation in each of $T_{\text{surf}}$, $T_{\text{tropo}}$, or $T_{\text{exo}}$ to examine the details of how each parameter affects $f$. \change{The full array}{This set} of temperature profiles is shown in Figure \ref{fig:alltemps}.

\subsubsection{Water Profiles} \label{sect:waterprofs}

\begin{figure}
	\centering
	\includegraphics[width=0.75\linewidth]{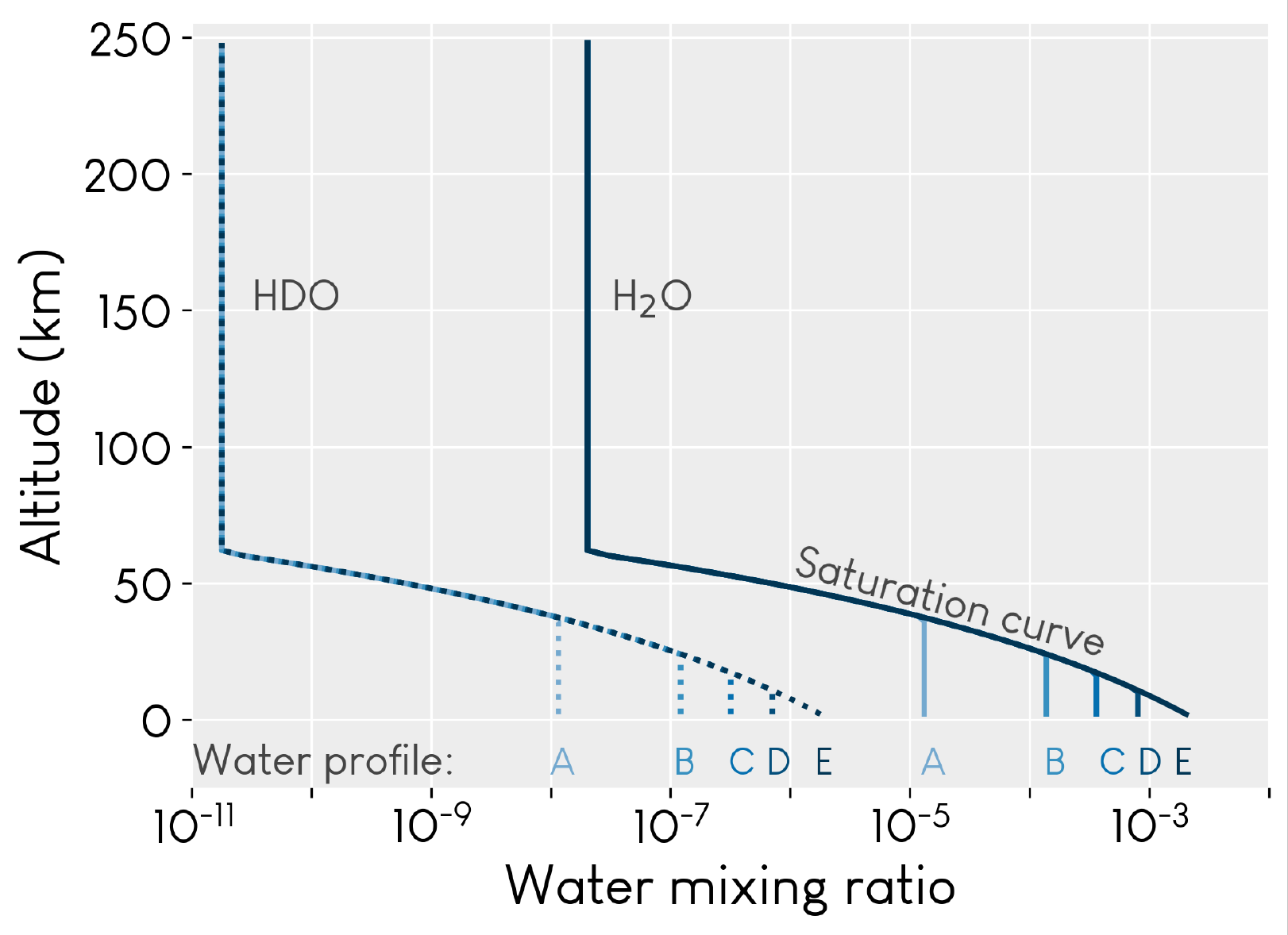}
	\caption{Water vapor profiles used in our model. A single profile, e.g. A, comprises both H$_2$O (solid lines) and HDO (dotted). Profiles are constrained by requiring that [H$_2$O]+[HDO] = 1 pr $\mu$m (profile A), 10 (B), 25 (C), 50 (D), or 100 (E) and that the HDO profile is equal to 5.5 $\times$ SMOW $\times$ the H$_2$O profile. Profiles differ in the well-mixed lower atmosphere and are the same once they reach the saturation vapor pressure curve. Water vapor in the mesosphere and upper atmosphere is negligible on average over long time scales, like those we model, although it may change on short time scales (see text). Profile B (10 pr $\mu$m) is used for our standard atmosphere.}
	\label{fig:water_profiles}
\end{figure}

H$_2$O and HDO profiles used in the model are shown in Figure \ref{fig:water_profiles}. We require that the profiles have total water content (H$_2$O + HDO) equal to 1, 10, 25, 50, or 100 pr $\mu$m (precipitable micrometers), with H$_2$O making up most of the share. Higher concentrations of water vapor would require a supersaturated atmosphere; while there is observational evidence of supersaturation at upper altitudes in specific cases\remove{,} \cite{Maltagliati2011, Fedorova2020}, our model does not include it. We use the 10 pr $\mu$m profile to represent the long-term standard atmosphere, a value in agreement with observations \cite{Lammer2003, Smith2004}, although more recent observations \cite{Heavens2018, Vandaele2019} and modeling \cite{Shaposhnikov2019} suggest that local water vapor concentrations can reach higher values, up to 150 pr $\mu$m, on very short timescales, particularly during dust storms. We assume that the lower atmosphere is well-mixed, such that the water vapor mixing ratio is constant. At the hygropause, usually between 25 and 50 km \cite{Kras2000, Heavens2018}, water begins to condense, and its mixing ratio follows the saturation vapor pressure curve until it becomes negligible in the upper atmosphere \cite{Heavens2018}. Although HDO preferentially condenses compared to H$_2$O \cite{Montmessin2005}, it never approaches saturation in our model atmosphere, allowing us to use the same empirical saturation vapor pressure equation \cite{Marti1993} for both H$_2$O and HDO. This is helpful, as no empirical equation for HDO exists, and the enthalpies of HDO under Mars-like conditions are very sparsely studied.

Although observations \cite{Villanueva2015} and modeling \cite{Fouchet1999, Bertaux2001} have shown that atmospheric D/H varies between 1-10$\times$ SMOW depending on the species it is measured in, altitude, and latitude/longitude, we tested these variations and determined that they had no effect on our results. We therefore multiply the initial profiles of H-bearing species by the D/H ratio of 5.5$\times$ SMOW to create the D-bearing profiles. The number densities of H$_2$O and HDO remain fixed during the simulation to represent the standard water abundance, though they are used to calculate chemical reaction rates.

\section{Results: \change{Non-thermal Escape Critical to Understanding the Fractionation Factor}{Temperature Variations and Non-thermal Escape Critical to Understanding the Fractionation Factor}} \label{sect:results}

\begin{figure}
	\centering
	\includegraphics[width=1\linewidth]{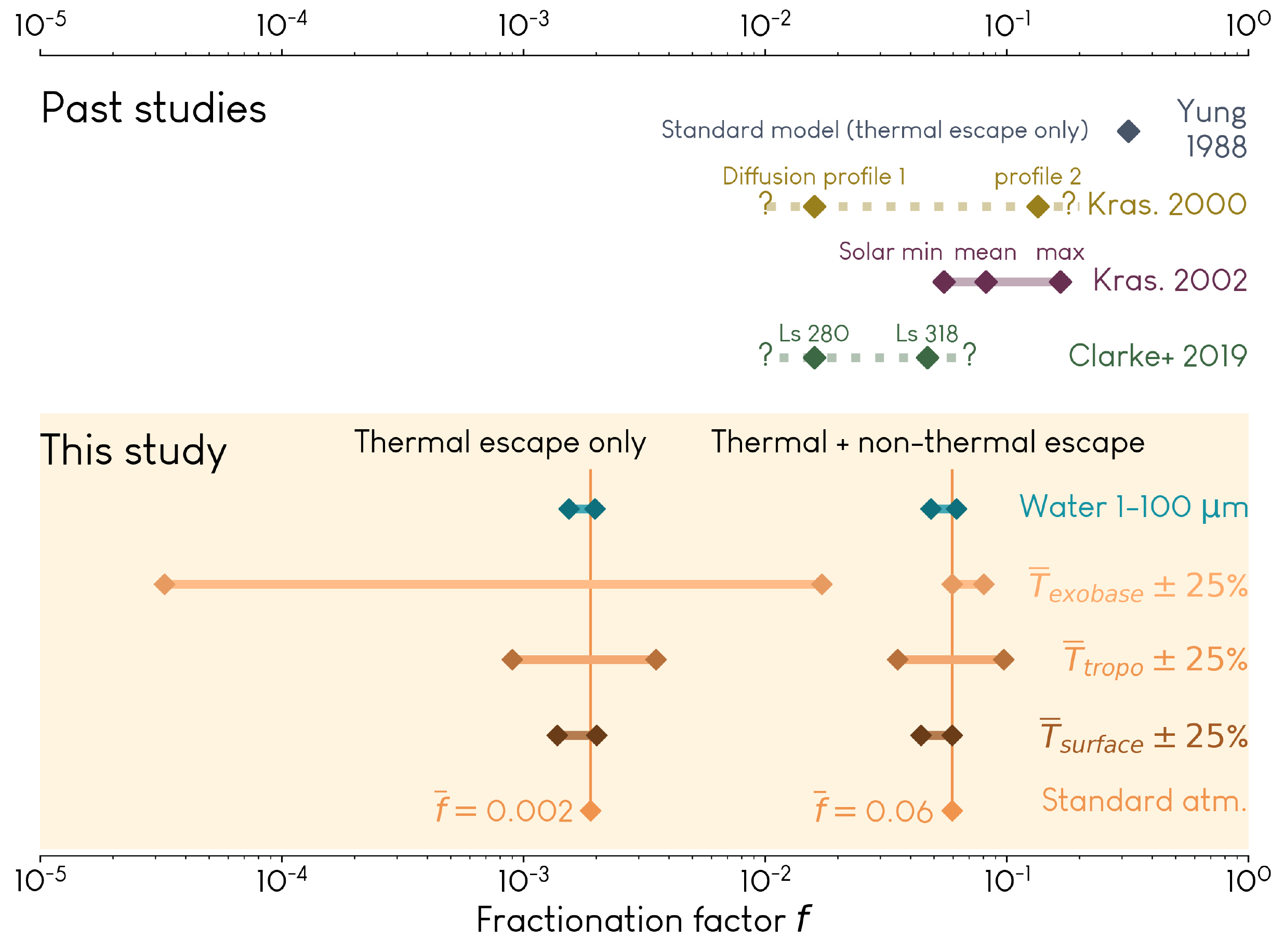}
	\caption{Results for the fractionation factor from this study (lower panel) and in past studies (upper panel). Bars represent the approximate range. Dotted lines with question marks indicate a study where the cases chosen did not necessarily represent end-member cases, so the true range is uncertain. Details of the dependence of $f$ on temperatures and water vapor (orange and blue bars in lower panel) are shown in Figures \ref{fig:f-vs-temps} and \ref{fig:water_plots_combined}. A numerical table of our results is available in Table S5.}
	\label{fig:f-results-plot}
\end{figure}

Figure \ref{fig:f-results-plot} shows the range of the fractionation factor as a function of each temperature and water vapor parameter, using the temperature profiles in Figure \ref{fig:temp_profiles} and the water vapor profiles in Figure \ref{fig:water_profiles}\remove{--that is, the standard profiles and the plausible climate extrema profiles}. Results for the broad range of temperatures shown in Figure \ref{fig:alltemps} are discussed in Section \ref{sect:tempdeets}.

For thermal escape only, we find that \change{the fractionation factor}{$f$} is 1-3 orders of magnitude lower than the original value by \citeA{Yung1988}. The primary reason for this difference is the exobase temperature (they use 364 K, \add{while} we use a maximum of 250 K\add{, which is more consistent with modern measurements}). \change{Additionally, they allow their model to self-consistently solve for water vapor number density above 80 km, while our entire profile is fixed. Updates in chemical and photochemical reaction rates over the last three decades are the last key difference.}{Other minor differences are as described in Section \ref{sect:repro}} Details of the dependence of $f$ on each parameter are discussed in sections \ref{sect:tempdeets} and \ref{sect:waterdeets}.

Because our model does not include an ionosphere, we do not model the effects of non-thermal escape processes\change{, including }{ (e.g. }sputtering, photochemical escape\change{, ion outflow, ion pickup, or bulk ion escape}{)}\add{, and only model thermal escape. This makes it difficult to compare with other studies which do include non-thermal escape (e.g. \citeNP{Kras2002})}. In order to \add{compare with that study, we must estimate $v_{nt}$, the non-thermal escape velocity, which is not part of our model. To do so,} \remove{approximate the effect of non-thermal escape, }we calculated the ratio of thermal \remove{($v_t$) }to non-thermal \remove{($v_{nt}$) }effusion velocities \add{($r=v_t/v_{nt}$)} for \remove{the }H, H$_2$, D, and HD \remove{species }in the model used by \citeA{Kras2002}. \change{We then used our model results for $v_t$ and}{We then divided our $v_t$ by} the ratio \add{$r$} to \add{get an }estimate \add{of} non-thermal effusion velocities \change{for our modeled temperatures. This allowed}{$v_{nt}$ at the temperatures modeled by \citeA{Kras2002}: 200 K, 270 K, and 350 K. We extrapolated this estimate down to 150 K, the lowest temperature in our model, by fitting a 2nd order polynomial to the estimates for $v_{nt}$, allowing} us to estimate the role that non-thermal escape plays in setting $f$\change{. T}{ for the temperature profiles shown in Figure \ref{fig:temp_profiles}. Though the estimation method is imperfect, it provides a rough estimation while avoiding unphysical velocity values at low temperatures, and gives}\remove{the resulting} values of $f$ are consistent with \citeA{Kras2000} and \citeA{Kras2002}, as well as more recent observations using MAVEN/IUVS \cite{Clarke2019}. Notably, our highest value of $f$ is approximately a factor of 3 larger than the lowest, in agreement with \citeA{Kras2002}. \add{In future work beyond the scope of this paper, we plan to directly model non-thermal contributions, enabling a better model comparison.}

\begin{figure}
	\centering
	\includegraphics[width=1\linewidth]{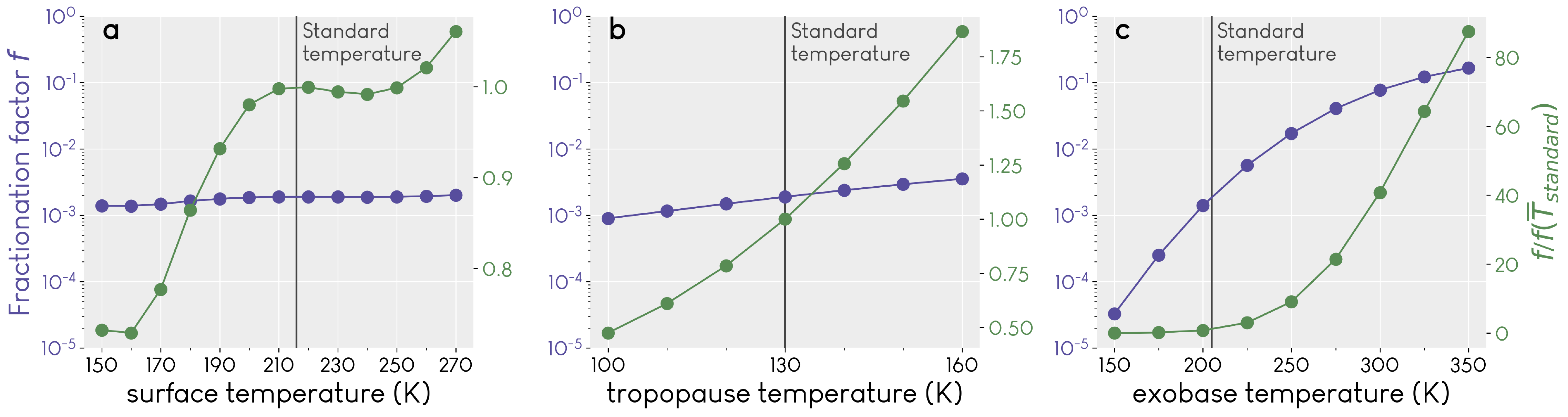}
	\caption{Dependence of the fractionation factor $f$ on changes in the surface, tropopause, and exobase temperatures. The standard value of each is marked by a black vertical line. The left (purple) axis shows the value of $f$, while the right (green) axis shows the relative change of $f$ with respect to that calculated for the standard temperature.}
	\label{fig:f-vs-temps}
\end{figure}

\subsection{Fractionation Factor \add{for Thermal Escape }Strongly Controlled by Exobase Temperature\remove{ in Thermal Escape}} \label{sect:tempdeets}

\begin{figure}
	\centering
	\includegraphics[width=1\linewidth]{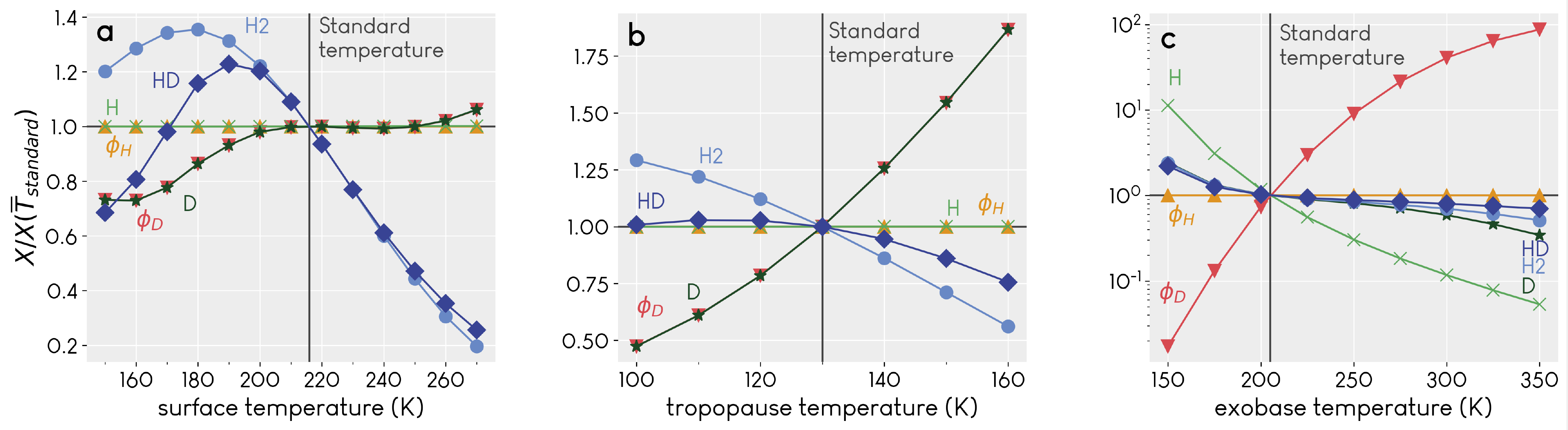}
	\caption{Change in exobase abundances of H- and D-bearing species or escape fluxes ($\phi$) as a function of temperature for thermal escape only. \change{$\phi_{\text{H}}$ includes loss from H, H$_2$, and HD, while $\phi_{\text{D}}$ includes loss via D and HD.}{$\phi_{\text{H}}$ and $\phi_{\text{D}}$ are calculated as in equations \ref{eq:escfluxH} and \ref{eq:escfluxD}.} In \add{panels }a and b, $\phi_{\text{D}}$ (and thus $f$ in Figure \ref{fig:f-vs-temps}a and b) closely tracks the abundance of atomic D. In panel c, changes in the abundance of H, D, H$_2$ and HD are caused by both escape to space and supply by diffusion from below. \change{Because of D's low abundance, }{D is not diffusion limited, so} $\phi_{\text{D}}$ responds more strongly to temperature forcing than H. Note the linear y-scale in panels a and b and the log scale in panel c. \add{In all panels, $\phi_{\text{H}}$ is on the order of $\sim$$10^8$ cm$^{-2}$s$^{-1}$, but differences between simulations are on the order of 10, making the relative variation of $\phi_{\text{H}}$ $\approxeq 1$.}}
	\label{fig:temps_Hplot}
\end{figure}

Figure \ref{fig:f-vs-temps} shows in detail how $f$ varies with each temperature parameter \add{for the modeled thermal escape only}. \change{In these cases, we only report results for modeled thermal escape,}{This approach allows us} \remove{in order }to focus on what we can learn about $f$ from our model, and refrain from drawing any strong conclusions about \change{what}{specific} effects \remove{may be} introduced by non-thermal escape before we can fully model it. \add{The results show that an increase in temperature in any part of the atmosphere leads to an increase of $f$ (less fractionation). The effect is small when the temperature increase occurs in the lower atmosphere, and dramatic when the change occurs at the exobase. Understanding this behavior requires examining the temperature-dependent behavior of H, D, H$_2$, and HD abundances and the escape fluxes $\phi_{\text{D}}$ and $\phi_{\text{H}}$. This information is shown in Figure \ref{fig:temps_Hplot}, where we plot these values for each simulation, normalized to the simulation with the standard atmosphere.}

\remove{Though the effect is small, $f$ increases as a function of surface and tropopause temperature. The cause of this increase is revealed by examining how the absolute abundances of H, D, H$_2$, HD, and the escape fluxes $\phi_{\text{D}}$ and $\phi_{\text{H}}$ vary with each temperature parameter; this information is shown in Figure \ref{fig:temps_Hplot}. To visualize this, we calculate the ratio of these abundances and fluxes in a given simulation (e.g., $T_{\text{surf}}=190$ K) to the standard atmosphere simulation ($T_{\text{surf}}=216$ K). The standard atmosphere case thus has a ratio of 1, and any simulation in which a species abundance or flux increases (decreases) relative to the standard atmosphere will have a ratio greater than (less than) 1. }As a function of both surface and tropopause temperature, $\phi_{\text{D}}$ most closely tracks the abundance of atomic D at the exobase. \add{Per equation \ref{eq:f}, }$f$ depends directly on $\phi_{\text{D}}$, inversely on $\phi_{\text{H}}$, and inversely on $R_{dh,s}$. \remove{Because $R_{dh,s}$ never changes, and because}\add{$R_{dh,s}$ is constant, and any reduction of }$\phi_{\text{H}}$ \change{is consistent across all temperatures}{is offset by an increase in $\phi_{\text{D}}$ because stoichiometric balance requires that $\phi_{\text{H}}+\phi_{\text{D}}=2\phi_O$. This change is many orders of magnitude smaller than $\phi_{\text{H}}$ and thus not visible in Figure \ref{fig:temps_Hplot}, but comparable to $\phi_{\text{D}}$, making it easily visible.} \change{the increase of $f$ with surface or tropopause temperature is due to a preferential increase in D at the exobase due to chemical or photochemical reactions. The increase is not likely due to transport, as D is less able to diffuse upward.}{The increase of $f$ with $T_{\text{tropo}}$ is thus primarily because the higher temperatures in the mesosphere and thermosphere enable greater transport of D upwards. This response is possible because D is not diffusion-limited. Moderate increases in $\phi_{\text{D}}$ are a by-product of this effect (see Figure S6), as the escaping population is proportional to the upper-atmospheric abundance of D. Transport into the mesosphere may also be slightly enhanced by increased $T_{\text{surf}}$, but the main role of $T_{\text{surf}}$ is to drive the chemical reaction rates in the dense lower atmosphere, where they dominate over transport in controlling species abundances.}

In contrast, the exobase temperature \add{$T_{\text{exo}}$} has a far greater effect on the value of $f$, with values ranging from $10^{-5}$ to $10^{-1}$. This is unsurprising, as $f$ directly depends on \remove{the escape fluxes }$\phi_{\text{D}}$\remove{, $\phi_{\text{H}}$ at the exobase}.\remove{The escape flux is the product of the number density $n_X$ and the escape velocity, $v_{esc}$.} Because \remove{the }thermal population\add{s} \change{of H is}{are} assumed to be Maxwellian, we take the \remove{escape }velocity \add{in equations \ref{eq:escfluxH} and \ref{eq:escfluxD} }to be the effusion velocity, which \change{directly depends}{depends directly} on \change{the temperature of the exobase}{$T_{\text{exo}}$}. \change{D is preferentially affected compared to H; in Figure \ref{fig:temps_Hplot}c, a much larger decrease in the abundance of H at the exobase compared to D is revealed, leading to a relative increase in $\phi_{\text{D}}$ compared to $\phi_{\text{H}}$ and an increase of $f$. This is likely due to greater diffusive separation of H in the heterosphere at low temperature.}{As $T_{\text{exo}}$ rises, the warmer thermosphere to enable enhanced vertical transport of D. Escape of D is enhanced enough that $\phi_{\text{D}}$ no longer closely tracks the abundance of atomic D. To compensate for the increased loss of D while maintaining stoichiometric balance, $\phi_{\text{H}}$ must decrease by a negligible amount (see Figure S7).}

\begin{figure}
	\centering
	\includegraphics[width=0.75\linewidth]{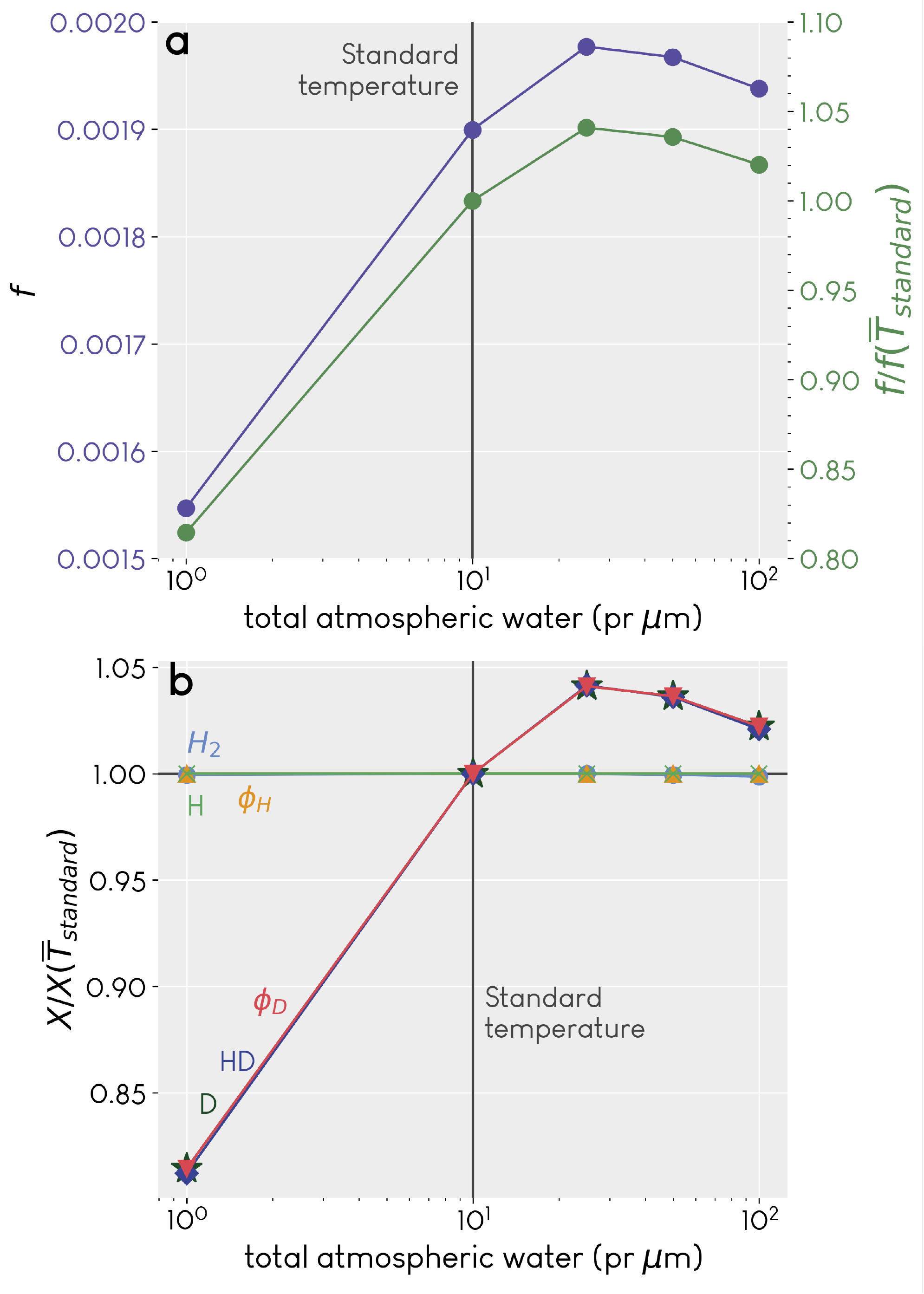}
	\caption{a) Fractionation factor \add{$f$ }as a function of water vapor column abundance, shown for concentrations of 1, 10, 25, 50, and 100 pr $\mathrm{\mu}$m, for thermal escape only. b) Same as Figure \ref{fig:temps_Hplot}, but as a function of water vapor. Here, $\phi_{\text{D}}$ and $f$ track the abundances of both D and HD.}
	\label{fig:water_plots_combined}
\end{figure}

\subsection{Fractionation Factor Depends Weakly on Water Vapor Column Abundance} \label{sect:waterdeets}

The fractionation factor as a function of total water vapor is shown in Figure \ref{fig:water_plots_combined}a, and the comparison of abundances and fluxes of H- and D-bearing species in Figure \ref{fig:water_plots_combined}b. As in the previous section, the increase of $f$ with additional water vapor is correlated with an increased abundance of D at the exobase, but also HD. The total water vapor has little effect on $f$, likely because the absolute abundance of water changes neither the D/H ratio in water or the processes by which it is fractionated. The small variation with respect to water vapor thus reflects the influence of minor differences in H$_2$O and HDO chemical and photochemical reactions. In order to more fully characterize the effects of water vapor on \change{the fractionation factor}{$f$}, the model will have to be modified to allow variable water vapor profiles.

\begin{figure}
	\centering
	\includegraphics[width=1\linewidth]{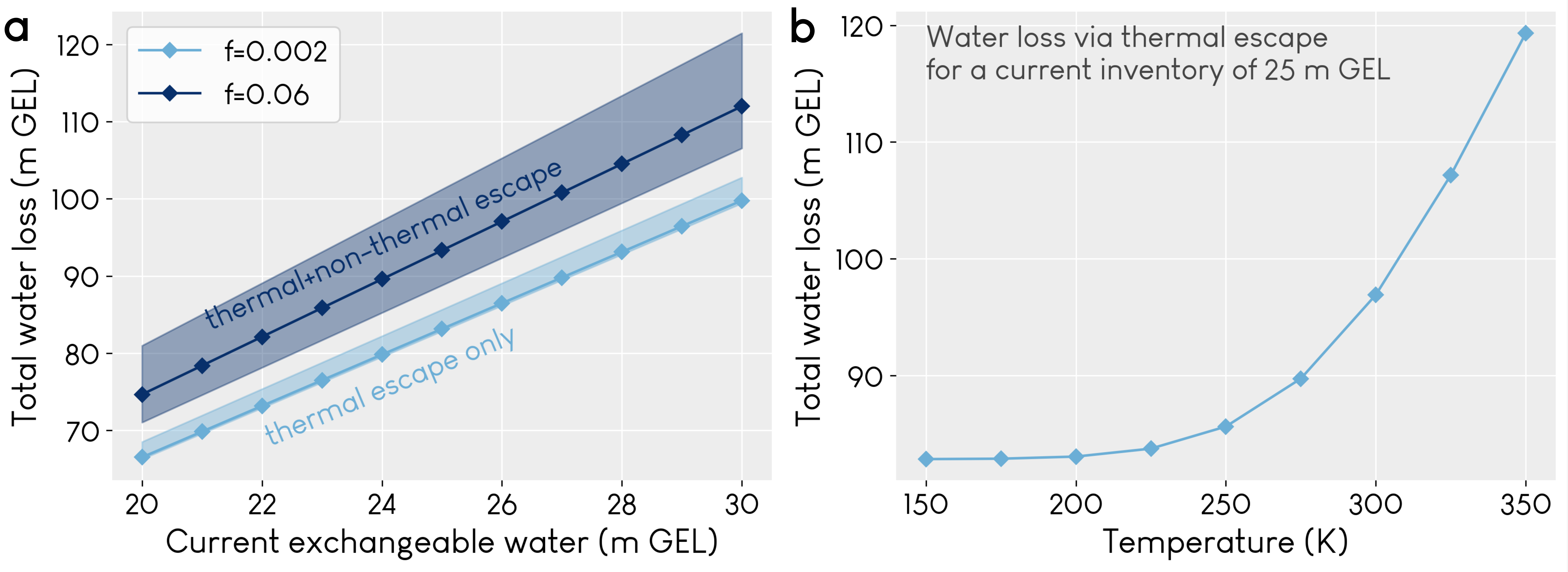}
	\caption{Water lost from Mars \add{a)} as a function of the current exchangeable water budget and the fractionation factor\add{ for the standard atmosphere ($f$$=$$0.002$ for thermal escape and $f$$=$$0.06$ for thermal and non-thermal together). The shaded regions represent the extrema of water loss, calculated for the extrema of $f$ of each escape type from our results. The lower bound for thermal escape is close to that of the standard case because water loss is insensitive to $f$ for $f$$<$$0.01$. b) as a function of exobase temperature and $f$ for that temperature as calculated in the model. Water loss is} calculated using Equation \ref{eq:rd}\remove{, where the slope of each line is $\left( R_{dh}(t)/R_{dh}(0) \right)^{1/(1-f)} - 1$}. \change{We use $R_{dh}(t) = 5.5 \times$ SMOW, $R_{dh}(0)= 1.275 \times$ SMOW \cite{Villanueva2015}.}{Equation parameters are described in Section \ref{sect:mapping}.} \add{Because we do not model non-thermal escape, the top line in panel a includes an approximation of the effect of non-thermal escape as described in the main text.} \remove{For thermal escape only, we use our result for the standard atmosphere, $f=0.002$; for the thermal and non-thermal case, $f=0.06$. The shaded regions represent the extrema of water loss, calculated for the extrema of $f$ of each escape type from our results. The lower bound for thermal escape is close to that of the standard case because water loss is insensitive to $f$ for $f<0.01$.}}
	\label{fig:waterloss}
\end{figure}

\subsection{Mapping Fractionation Factor Results to Integrated Water Loss} \label{sect:mapping}

We can determine the magnitude of water loss on Mars by using our results for $f$ as input to Equation \ref{eq:rd}. These results are shown in Figure \ref{fig:waterloss}. In order to use Equation \ref{eq:rd} to plot past water loss, we must set values for the current water inventory $W(t)$, the current D/H ratio $R_{dh}(t)$, and the ancient Martian D/H ratio, $R_{dh}(0)$.

For $W(t)$, we use the \add{aforementioned} range 20-30 m GEL \change{to encompass the range of observations of}{for} the current exchangeable water budget of Mars \remove{\cite{Villanueva2015, Lasue2013}}. Exchangeable water is water that is able to move between surface deposits and the atmosphere; its D/H ratio increases due to escape to space. Non-exchangeable water, being unaffected by escape to space, would have its original D/H value.

For $R_{dh}(0)$, we follow \add{\citeA{Kurokawa2014} and} \citeA{Villanueva2015} and use 1.275 $\times$ SMOW, in agreement with the measurement of D/H in the 4.5 billion year old melt inclusions in the Martian meteorite Yamato 980459 \cite{Usui2012}. Finally, we use 5.5 $\times$ SMOW for $R_{dh}(t)$.

Using these values, we calculate the \add{cumulative }water lost\remove{ over 4.5 billion years (Ga)} to be between about 66 and \change{123}{122} m GEL, depending on escape type and value of $f$. We compare these results with other estimates in the literature in the next section.

\section{Discussion} \label{sect:discussion}

Because \change{the fractionation factor}{$f$} depends directly on the escape fluxes $\phi_{\text{D}}$ and $\phi_{\text{H}}$, it is reasonable that the exobase temperature would most strongly affect $f$ \add{for thermal escape}. Disturbances in the lower atmosphere that may otherwise affect $f$ will be \change{generally depleted}{reduced} in amplitude by the time they propagate to the upper atmosphere. A larger $f$ at higher exobase temperatures also makes sense in the context of past work; the Mariner missions measured the exobase temperature to be 350 $\pm$ 100 K \cite{Anderson1971}, and \citeA{Yung1988} used $T_{\text{exo}}=364$ K to obtain $f=0.32$ for thermal escape only. However, these original Mariner measurements were highly uncertain; more recent data (discussed previously) indicate that $T_{\text{exo}}$ during solar mean and minimum is cold enough that $f$ for thermal escape is substantially smaller, and that non-thermal escape of D is critical to an accurate calculation of $f$.

\add{The importance of $T_{\text{tropo}}$ is also worth some discussion. In Figure \ref{fig:f-results-plot}, $T_{\text{tropo}}$ appears to be the parameter with the greatest control over $f$ for thermal + non-thermal escape together. It should be noted, however, that our estimates of non-thermal escape are approximate and extrapolated below 200 K, as described in Section \ref{sect:results}. For the simulations that varied $T_{\text{surf}}$ or $T_{\text{tropo}}$, each estimation of non-thermal escape was made assuming a constant value of $T_{\text{exo}}$ (205 K). On the other hand, different values of $T_{\text{exo}}$ were necessarily used in the simulations which tested the effects of varying it. This means that $f_{\text{thermal+nonthermal}}$ and $f_{\text{thermal}}$ as a function of $T_{\text{tropo}}$ appear to differ by a constant, whereas $f_{\text{thermal+nonthermal}}$ as a function of $T_{\text{exo}}$ may be artificially large at temperatures below 200 K (see Figure S5). Despite these uncertainties, $T_{\text{tropo}}$ is still important to the value of  $f$, as increased $T_{\text{tropo}}$ also increases the temperature of the thermosphere, enabling greater upward transport and contributing to overall escape (see Figure S6).}

The relationship of $\phi_{\text{D}}$ \add{from thermal escape} to the abundances of atomic D and HD is not immediately obvious. In Figure \ref{fig:temps_Hplot}a and b, $\phi_{\text{D}}$ most closely tracks the abundance of atomic D at the exobase because it is much more abundant than HD. In all of the simulations represented in these panels, the exobase temperature is 205 K, a value too low for escape of HD to contribute significantly to D loss. Only at high exobase temperatures (Figure \ref{fig:temps_Hplot}c) or high concentrations of water \remove{near the exobase }(Figure \ref{fig:water_plots_combined}b) does \change{the HD line get closer to the $\phi_{\text{D}}$ line}{$\phi_{\text{D}}$ appear to track the HD abundance}, indicating HD is abundant enough to contribute more to D loss. In general, in Figures \ref{fig:temps_Hplot} and \ref{fig:water_plots_combined}b, \change{the more closely the line tracks either the D or HD lines, the more abundant that species is at the exobase. A higher abundance leads to a greater contribution to escape}{an increase in $\phi_{\text{D}}$ is correlated with an increase in the abundance of D, except when $T_{\text{exo}}$ increases and escape is dramatically enhanced. More abundant deuterium means more deuterium available to escape}; in most cases, loss of \change{D}{deuterium} (\change{H}{hydrogen}) via \change{the atomic form}{D (H)} dominates, but at high exobase temperatures, loss via the molecular form HD (H$_2$) can reach higher values, up to 5\% (20\%), as shown in Figure S4.

\begin{figure}
	\centering
	\includegraphics[width=1\linewidth]{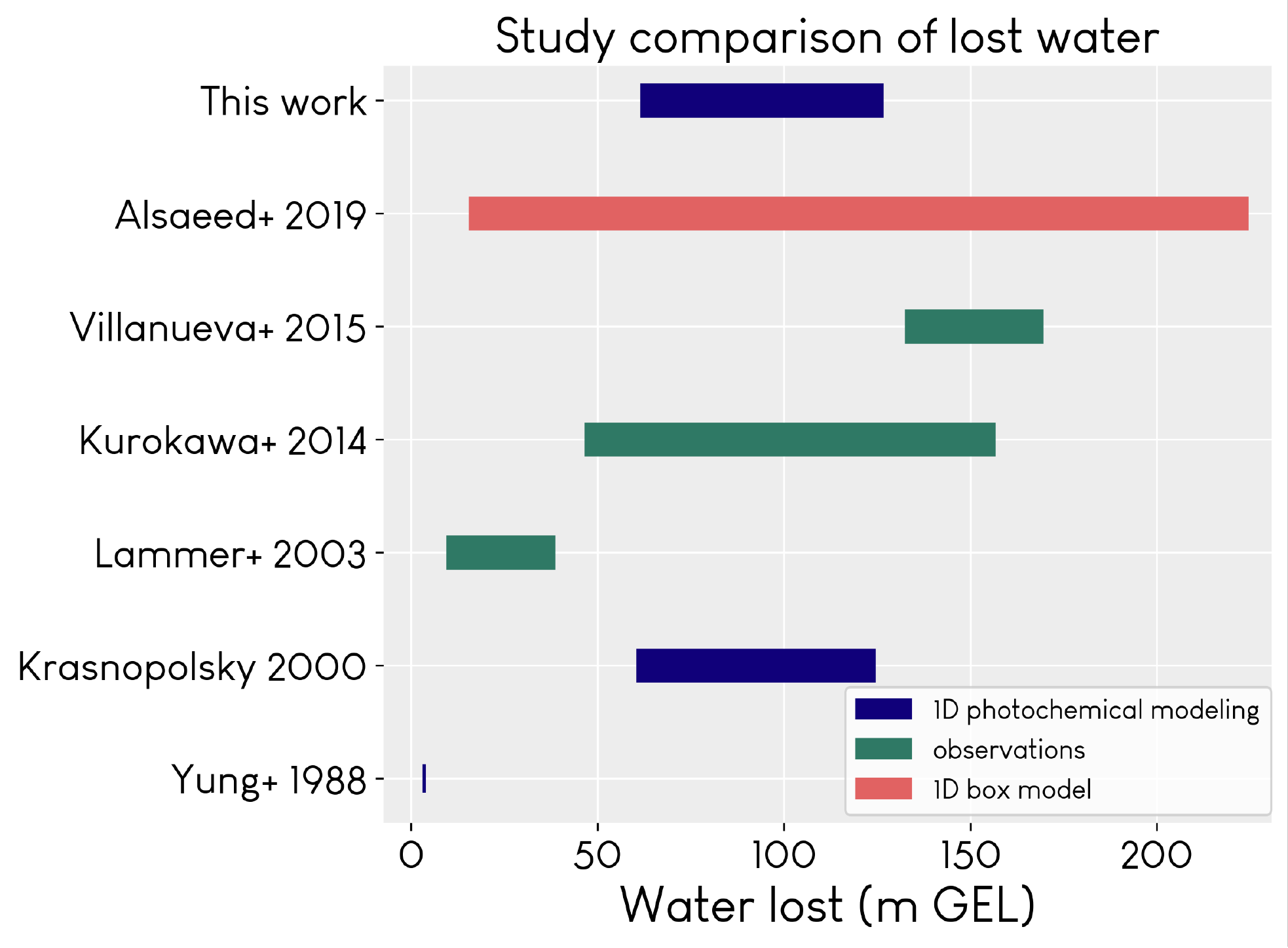}
	\caption{Estimates of water lost from Mars by various studies. }
	\label{fig:water_loss_comparison}
\end{figure}

A comparison of our results for water loss to those of other similar studies is shown in Figure \ref{fig:water_loss_comparison}. Overall, our results agree reasonably well with these other studies. Our results are a little lower than those by \citeA{Villanueva2015}, who assume a higher atmospheric D/H ratio (7-8 $\times$ SMOW), and a little higher than \citeA{Lammer2003}, who \change{use both}{assume} a higher \remove{assumed }D/H ratio for early Mars (1.2-2.6 $\times$ SMOW) and a lower estimate of the current exchangeable water (3.3-15 m GEL). The original study by \citeA{Yung1988} is an outlier in this case because they were attempting to determine both the current water inventory and the amount lost, and did not have the benefit of the many Mars missions and observations that we have today. \add{The recent work by \citeA{Alsaeed2019} is unique compared to the other studies in this figure in that they allow water to be added to the atmosphere via volcanic outgassing, so that their results represent a large possible solution space and are less directly comparable than the other studies.}

Our results for water loss also bring up an important point with regard to escape rates. It is common when estimating water loss on Mars to assume that the escape fluxes $\phi_{\text{H}}$ and $\phi_{\text{D}}$ are constant and that the water inventory decreases linearly with time. This is an often necessary but imperfect assumption due to the many unknowns involved, including historical rates of atmospheric escape and their evolution in light of Mars' chaotically evolving obliquity. Assuming linear loss with time (and neglecting $\phi_{\text{D}}$, which is far slower than $\phi_{\text{H}}$) gives $\phi_{\text{H}} = W_{lost} / t$, where $t$ is the time over which the water has been lost. Using our results for water loss, even the smallest amount lost (about 60 m GEL) requires an escape rate of \remove{approximately} $\sim3 \times 10^{9}$ cm$^{-2}$ s$^{-1}$, an order of magnitude higher than what we currently observe for escape rates of H from Mars \cite{Jakosky2018}\remove{and find in our modeling, in which $\phi_+ \phi_{\text{D}} = 2\phi_O$}. This is an indication that escape rates were likely higher in the past due to a variety of factors, especially in the context of a more UV-active young sun \cite{Jakosky2018}, or that surface interactions play a larger role that has \add{been previously explored--e.g., with regard to oxygen deposition \cite{Zahnle2008}--but} not yet \remove{been }fully quantified.

\begin{figure}
	\centering
	\includegraphics[width=1\linewidth]{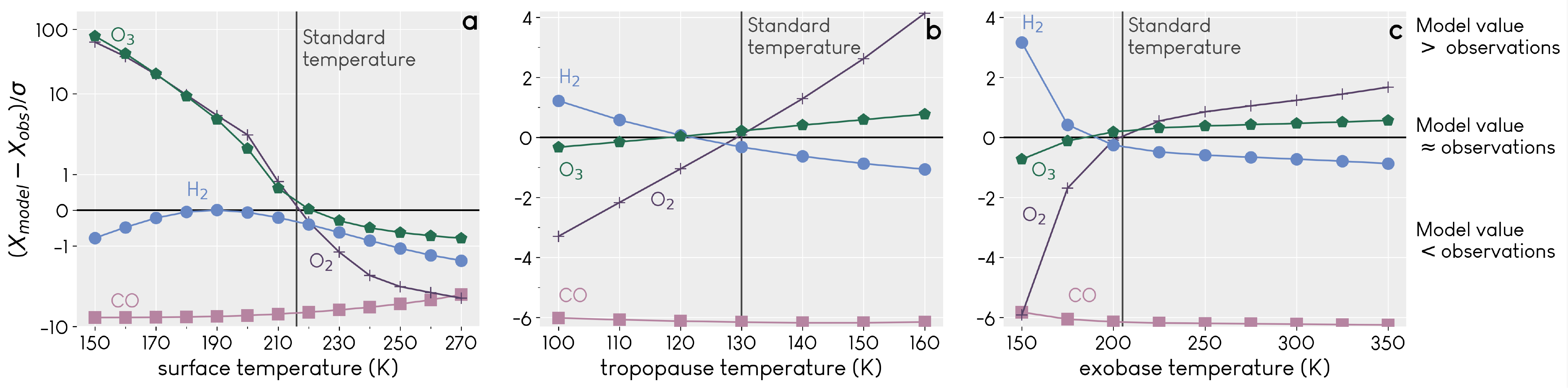}
	\caption{Comparison of model output values to measured values as a means of determining appropriateness of our temperature assumptions. See text for measurement citations. O$_3$ is measured in $\mu$m-atm. O$_2$ and CO are measured as the mixing ratio at the surface. H$_2$ is measured with the total abundance in ppm in the lower atmosphere (0-80 km). The y-axis is the difference between model output and measurement, weighted by the uncertainty in the measurement; the closer a point is to the 0 line, the more similar the model output and measurement.}
	\label{fig:output_vs_data_temps}
\end{figure}

\begin{figure}
	\centering
	\includegraphics[width=0.5\linewidth]{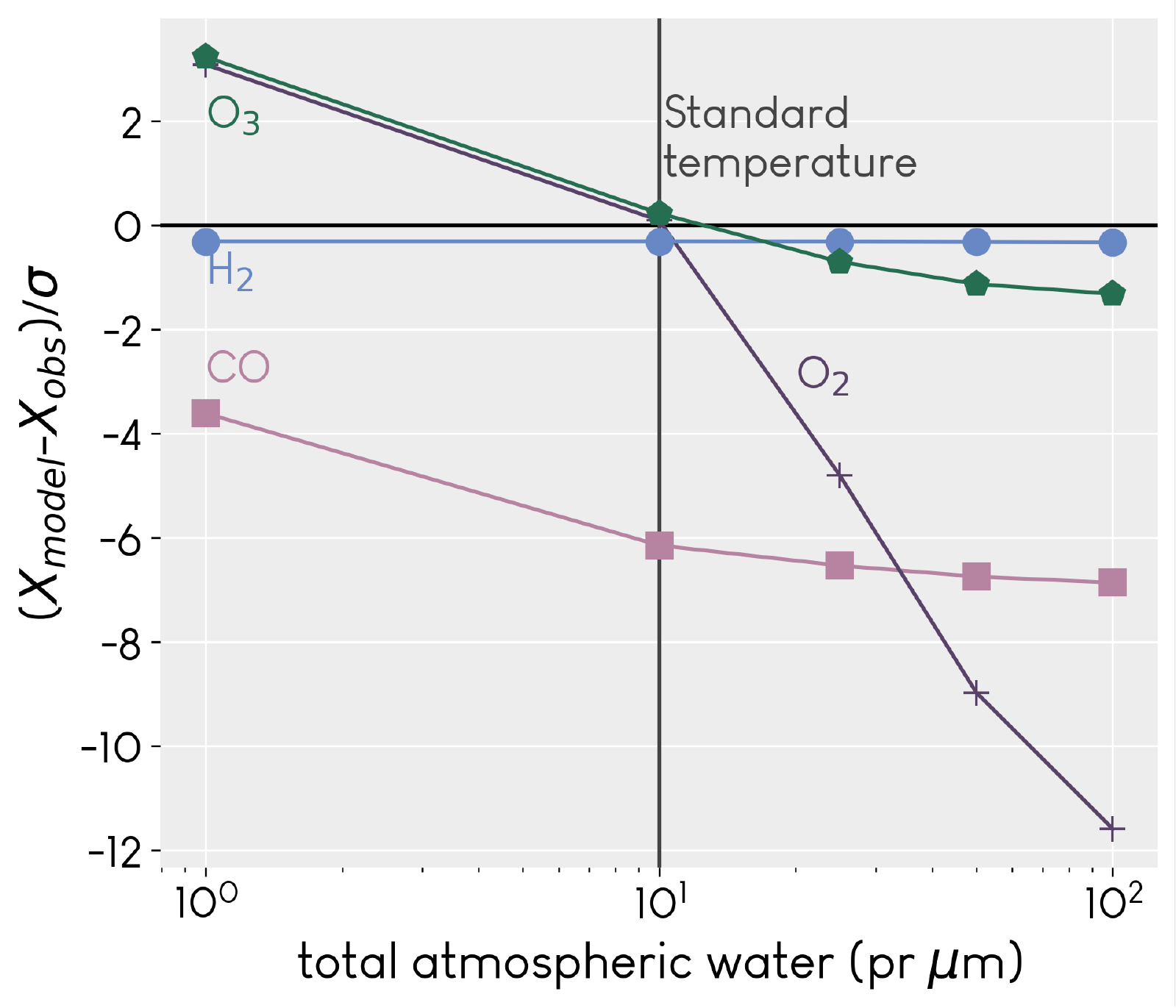}
	\caption{The same as Figure \ref{fig:output_vs_data_temps}, but for model runs where we varied the water vapor content of the atmosphere.}
	\label{fig:output_vs_data_water}
\end{figure}

As a way to gain insight about our results, we compared the concentrations of a few molecular species in our model with available measurements (Figures \ref{fig:output_vs_data_temps} and \ref{fig:output_vs_data_water}). The measurements we used were the inferred lower atmospheric abundance of H$_2=$ 15 $\pm$ 5 ppm \cite{Kras2001}; a global mean O$_3$ abundance of 1.2 $\mu$m-atm, extracted from maps by \citeA{Clancy2016}; and mixing ratios for O$_2$ and CO at the surface equal to $(5.8 \pm 0.8)\times 10^{-4}$ and $(1.61 \pm 0.09)\times 10^{-3}$ \cite{Trainer2019}. These comparisons indicate the model conditions which may be more similar or dissimilar to the current state of Mars. As one example, model results that used a particularly low temperature as input (for example, models with $T_{\text{surf}}<190$ or $T_{\text{exo}}<175$) diverge greatly from measurements of all molecular species. These model results thus represent a significant perturbation to the photochemical system as compared to modern Mars. It is also important to note that O$_3$ and O$_2$ are related, as O$_3$ is created and destroyed via interactions between O$_2$ and O. CO sticks out as an obvious problem; this is not surprising, as many photochemical models also have difficulty in reproducing the observed values \cite{Kras2010}. Some models come close (e.g. \citeA{Zahnle2008}), usually only when another parameter changes significantly. Our model also underestimates CO, reaffirming the ongoing need for study in this area. Apart from CO, the difference between our model and measurements is mostly small, indicating that the standard atmosphere we chose was reasonable.

\note{This is the paragraph that was moved.} Our results represent a peri-modern global scenario; $f$ has likely changed over time in ways that our model does not account for. In this work, we consider only the exchangeable reservoirs of water on Mars without including any type of surface deposition, which comprises multiple processes with potentially different fractionation factors. Fractionation may also vary on seasonal timescales, especially around the poles, as HDO preferentially condenses and may also have a different sublimation rate compared to H$_2$O. It has certainly varied over geological time scales. We run the model for 10 million years to equilibrium, though it would not necessarily have been in equilibrium throughout its 4.5 billion year history. This also means that atmospheric escape rates would not have been constant in time\change{. We assume escape rates to space to be constant}{, although we assume them to be} because their time evolution is unknown. Mars' chaotically evolving obliquity on time scales greater than 10 million years is a major reason for this lack of a definitive paleo-climate timeline. Characterization of escape rates through time is therefore a critical, but daunting, subject for future modeling efforts. On early Mars, \remove{$f$ would also have been different due to }the more UV-active young sun\remove{, which} would have enhanced non-thermal escape rates \cite{Jakosky2018}\change{.}{,} \add{allowing $f$ to grow larger due to the enhancement of $\phi_{\text{D}}$. This is allowed even in long-term chemical equilibrium; when no D is present, $\phi_{\text{H}}=2\phi_O$, but when D is present, the requirement instead becomes $\phi_{\text{H}}+\phi_{\text{D}}=2\phi_O$.} For all these reasons, we expect that our results for water loss are a lower bound.

\section{Conclusions}
Our results in Figure \ref{fig:f-results-plot} and Table S5 show that if only thermal escape is\add{ }considered, \add{the exobase temperature has the strongest effect on the fractionation factor, whereas when non-thermal escape is included, the temperatures of the tropopause and exobase become comparably important (given the uncertainty in our non-thermal escape estimation). While the exobase and upper atmosphere have been the focus of many recent studies, the Martian mesosphere is less well-studied, but is worthy of future modeling and observational efforts. The tropopause temperature affects mesospheric chemistry, which is especially important considering the seasonal transport of water to these altitudes \cite{Chaffin2017, Heavens2018, Shaposhnikov2019, Fedorova2020}. Because $T_{\text{tropo}}$ is the minimum temperature in the atmosphere, a larger value also implies a warmer thermosphere (see Figure \ref{fig:alltemps}), which may contribute to enhanced transport, especially above the homopause.

Our results also show how important non-thermal escape is to accurately calculate $f$. For thermal escape processes only, }D is almost completely retained on Mars compared to H. This is especially true near solar maximum, when most atmospheric \change{escape overall}{loss} occurs as thermal escape of H. During solar mean and minimum, however, thermal escape of \add{both D and} H is low\change{, and the fact that}{enough that }the non-thermal \remove{escape dominates }loss of D and HD \remove{\cite{Kras1998,Gacesa2012}} becomes much more significant \add{\cite{Kras1998,Gacesa2012}}. \change{Our analysis}{Figure \ref{fig:f-results-plot}} shows that including non-thermal escape \remove{significantly} increases $f$ by an order of magnitude or more for all atmospheric conditions\remove{, and that the tropopause temperature is the parameter with the greatest effect on $f$ (Figure \ref{fig:f-results-plot})}.  Studies of only thermal escape are therefore not likely to provide a reasonable estimate of $f$\add{. We therefore advise that future modeling studies that calculate $f$ should include non-thermal escape; this will also enable better comparison with recent observationally-derived values of $f$ (e.g. \citeNP{Clarke2019})}. \remove{

It is unclear whether the tropopause temperature's importance relates to a real, yet unknown, physical phenomenon, or whether it is an artifact resulting from our estimation of non-thermal escape. More modeling including non-thermal escape and observations of mesospheric phenomena are necessary to understand this effect in detail.}

\note{The paragraph that was previously here has been moved to the end of the Discussion section}

\add{In this study, we have neglected interaction with the planetary surface, which is certainly important due to the unknown D/H of surface ice and polar caps and fractionating effects on sublimation and deposition. }Future work to understand the fractionation factor and atmospheric escape will need to link cross-disciplinary knowledge of surface and atmospheric processes. The history of water on Mars cannot be fully understood by only considering one or the other; they are inextricably linked. A more thorough understanding of exchange between different water reservoirs on and under the surface and in the atmosphere, as well as the variables affecting all types of atmospheric escape and water loss, will be instrumental in forming a more complete picture of the fractionation factor, and by extension water loss, on Mars.

%%% End of body of article

%%%%%%%%%%%%%%%%%%%%%%%%%%%%%%%%%%%%%%%%%%%%%%%%%%%%%%%%%%%%%%%%
%
%  ACKNOWLEDGMENTS
%
% The acknowledgments must list:
%
% >>>>	A statement that indicates to the reader where the data
% 	supporting the conclusions can be obtained (for example, in the
% 	references, tables, supporting information, and other databases).
%
% 	All funding sources related to this work from all authors
%
% 	Any real or perceived financial conflicts of interests for any
%	author
%
% 	Other affiliations for any author that may be perceived as
% 	having a conflict of interest with respect to the results of this
% 	paper.
%
%
% It is also the appropriate place to thank colleagues and other contributors.
% AGU does not normally allow dedications.

\acknowledgments
%Enter acknowledgments, including your data availability statement, here.
This work was supported by MDAP grant \#NNX14AM20G. Additionally, this material is based upon work supported by the National Science Foundation Graduate Research Fellowship Program under Grant No. DGE 1650115. Any opinions, findings, and conclusions or recommendations expressed in this material are those of the author(s) and do not necessarily reflect the views of the National Science Foundation. The authors would like to thank B. Jakosky, N. Alsaeed, L. Wernicke, and D. Brain for ongoing collaboration, feedback, and support. \change{The model, information necessary to construct inputs, and generated output can be found on the lead author's GitHub account at \url{https://github.com/emcangi/dh_fractionation}.}{The model, input information, and generated output can be found on Zenodo \cite{Cangi2020}.}

%% ------------------------------------------------------------------------ %%
%% References and Citations

\bibliography{Cangi_2020_FF_journal_edits}

\end{document}